\documentclass[preprint, amsmath, amssymb, preprintnumbers, superscriptaddress, floatfix, showpacs, showkeys, aps, prl]{revtex4-1}

\usepackage{graphicx}
\usepackage{float}
\usepackage{color}
\usepackage{booktabs}
\usepackage{multirow}
\begin{document}

\title{Potential ring of Dirac nodes in a new polymorph of Ca$_3$P$_2$}

\author {Lilia S. Xie}
\affiliation{Department of Chemistry, Princeton University, Princeton, New Jersey 08544, USA.}
\author {Leslie M. Schoop}
\affiliation{Department of Chemistry, Princeton University, Princeton, New Jersey 08544, USA.}
\author {Elizabeth M. Seibel}
\affiliation{Department of Chemistry, Princeton University, Princeton, New Jersey 08544, USA.}
\author {Quinn D. Gibson}
\affiliation{Department of Chemistry, Princeton University, Princeton, New Jersey 08544, USA.}
\author {Weiwei Xie}
\affiliation{Department of Chemistry, Princeton University, Princeton, New Jersey 08544, USA.}
\author {Robert J. Cava}
\email{rcava@princeton.edu}
\affiliation{Department of Chemistry, Princeton University, Princeton, New Jersey 08544, USA.}

\date{\today}

\begin{abstract}

We report the crystal structure of a new polymorph of Ca$_3$P$_2$, and an analysis of its electronic structure. The crystal structure was determined through Rietveld refinements of powder synchrotron x-ray diffraction data. Ca$_3$P$_2$ is found to be a variant of the Mn$_5$Si$_3$ structure type, with a Ca ion deficiency compared to the ideal 5:3 stoichiometry to yield a charge-balanced compound. We also report the observation of a secondary phase, Ca$_5$P$_3$H, in which the Ca and P sites are fully occupied and the presence of interstitial hydride ions creates a closed-shell electron-precise compound. We show via electronic structure calculations of Ca$_3$P$_2$ that the compound is stabilized by a gap in the density of states compared to the hypothetical compound Ca$_5$P$_3$. Moreover, the calculated band structure of Ca$_3$P$_2$ indicates that it should be a three-dimensional Dirac semimetal with a highly unusual ring of Dirac nodes at the Fermi level. The Dirac states are protected against gap opening by a mirror plane in a manner analogous to graphene. The results suggest that further study of the electronic properties of Ca$_3$P$_2$ will be of interest.

\end{abstract}

\maketitle

\section{Introduction}

Through studies of the Ca--P chemical system, the compounds CaP, Ca$_3$P$_2$, CaP$_3$, CaP$_5$, and Ca$_5$P$_8$ have been reported to exist, but the full phase diagram is not known \cite{massalski1990binary,zaitsev1991thermodynamic,pytlewski1963study,dahlmann1973cap,franck1932contribution,hadenfeldt1994}. In particular, knowledge of the compound Ca$_3$P$_2$ is lacking. Ca$_3$P$_2$ is produced in industrial quantities and has been used in several applications, including as a rodenticide \cite{moran1988control}, an ingredient in pyrotechnic bombs \cite{faber1919history}, and a synthetic precursor to other phosphide compounds \cite{ripley1962preparation}. Computational studies have assumed that it crystallizes in the same structure type as Mg$_3$P$_2$ \cite{mokhtari2008density}, but to the best of the authors' knowledge, the full crystal structure Ca$_3$P$_2$ has never been reported. Furthermore, although the electronic characteristics of numerous semiconducting metal phosphide compounds such as InP \cite{hilsum1961semiconducting,duan2001indium}, GaP \cite{dean1966intrinsic,thomas1966isoelectronic}, and Cd$_3$P$_2$ \cite{haacke1964preparation},  have been characterized extensively, to date, there has been limited investigation of the electronic properties of calcium phosphides. The current study is motivated by the non-existence of a complete phase diagram for the Ca--P chemical system as well as the gap in our knowledge of the properties of calcium phosphides.

Here, we report the existence, crystal structure, and electronic structure analysis of a polymorph of Ca$_3$P$_2$ that crystallizes in the Mn$_5$Si$_3$ structure type (space group $P6_3/mcm$), distinct from a previously reported structure for Ca$_3$P$_2$ \cite{pytlewski1963study}. Rietveld refinements of high-resolution synchrotron x-ray diffraction data indicate that incomplete occupancy of the Ca sites yields the Ca$^{2+}$--P$^{3-}$ charge-balanced compound. We also report a new minority phase in the same space group corresponding to the formula Ca$_5$P$_3$H, in which interstitial hydride ions stabilize the full occupancy of the Ca sites by balancing the overall charge of the compound. Finally, we present electronic structure calculations predicting that Ca$_3$P$_2$ in the Mn$_5$Si$_3$ structure type is a three-dimensional (3D) Dirac semimetal with a highly unusual ring of electronic states with Dirac (i.e. linear $E$ vs. $k$) dispersions crossing the Fermi energy. This Dirac ring, comprising states in the $k_x$-$k_y$ plane, is protected by crystalline reflection symmetry and negligibly gapped by spin-orbit coupling in a manner highly analogous to the Dirac states of graphene \cite{chiu2014classification,CastroNeto2009}. Ca$_3$P$_2$ may therefore be a promising candidate for the further study of novel quantum transport phenomena in 3D systems.

\section{Experimental}

Calcium metal pieces (99.5\% purity) and red phosphorus powder (98.9\% purity) were combined in a stoichiometric ratio of 5:3 and sealed in a welded tantalum tube, with no additional purification measures undertaken. The tube was heated in an induction furnace for 2 hours at 1200 $^\circ$C as measured by a pyrometer. The resulting product was a fine blue-black powder that was sensitive to moisture. Commercial grade Ca$_3$P$_2$ was obtained from Alfa Aesar (CAS 1305-99-3) for comparison purposes. Initial structural investigation was carried out via powder x-ray diffraction (XRD) on a Bruker D8 Focus diffractometer with Cu K$\alpha$ radiation. Samples were placed in an airtight holder. 

High-resolution synchrotron XRD data was collected using the 11BM beamline at the Advanced Photon Source (APS) of Argonne National Laboratory using a wavelength of 0.41384 \AA~and a data collection range of 2$\theta$ = 0.5 degrees to 50 degrees. Samples were first loaded into 0.5 mm diameter glass capillaries and sealed in an argon atmosphere due to their air-sensitive nature. The capillaries were placed in 0.8 mm diameter Kapton tubes for measurement. The obtained XRD patterns were refined using the Rietveld method \cite{rietveld1969profile} and the program FullProf \cite{rodriguez1993fullprof}.

Electronic structure calculations were performed in the framework of density functional theory (DFT) using the \textsc{wien2k} \cite{blaha2001} code with a full-potential linearized augmented plane-wave and local orbitals [FP-LAPW + lo] basis \cite{singh2006,madsen2001,sjaestedt_alternative_2000} together with the Perdew-Becke-Ernzerhof (PBE) parameterization \cite{perdew_generalized_1996} of the Generalized Approximation (GGA) as the exchange-correlation functional. The plane wave cut-off parameter $R_{\text{MT}}K_{\text{MAX}}$ was set to 7 and the irreducible Brillouin zone was sampled by 133 $k$-points. The partial occupancy was modeled with the Virtual Crystal Approximation (VCA).

Crystal orbital Hamilton population (COHP) calculations was calculated via Tight-Binding Linear-Muffin-Tin-Orbital Atomic Sphere Approximation (TB-LMTO-ASA) using the Stuttgart code \cite{jepsen2000stuttgart,dronskowski1993crystal}. A mesh of $6 \times 6 \times 4$ $k$-points in the irreducible wedge of the first Brillouin zone was used to obtain COHP curves.

\section{Results and discussion}

\subsection{Structural refinement and composition of Ca$_3$P$_2$ and Ca$_5$P$_3$H}

Preliminary Rietveld refinements of powder XRD patterns taken on a laboratory diffractometer suggested that the obtained compound crystallized in the space group $P6_3/mcm$, analogous to the compound Ca$_5$As$_3$ \cite{Huetz1975}. These diffraction patterns were distinct from that of commercially available Ca$_3$P$_2$, which was amorphous.

\begin{figure}[htbp]
\centering
\includegraphics[width=12cm]{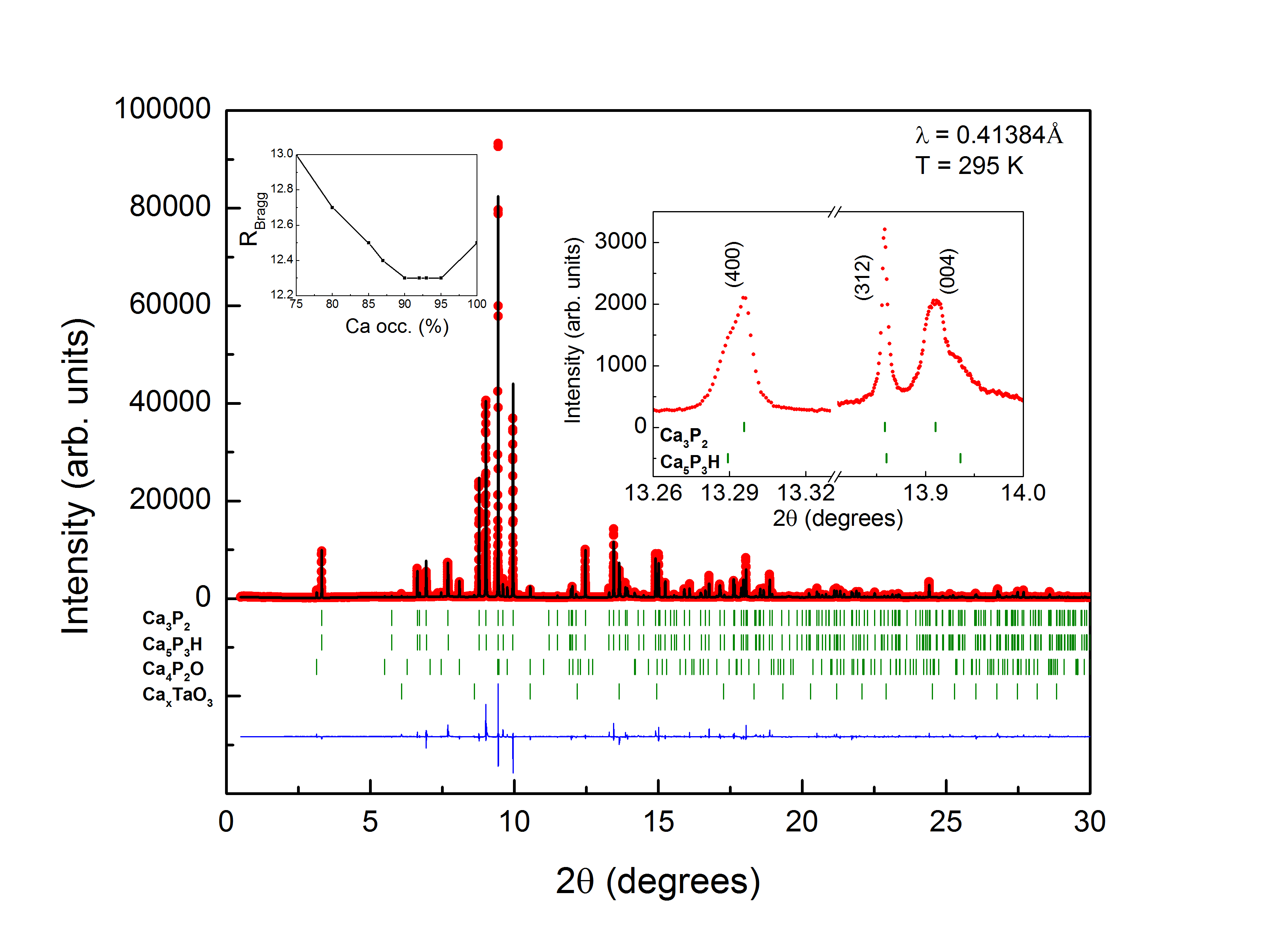}
\caption{Rietveld refinement of synchrotron XRD data. Peaks corresponding to the phases Ca$_3$P$_2$ and Ca$_5$P$_3$H, as well as Ca$_4$P$_2$O and Ca$_x$TaO$_3$ impurity phases are indicated. Left inset: $R_{\text{Bragg}}$ values for the majority phase refinement vs. Ca occupancy, wherein the lowest values of $R_{\text{Bragg}}$ correspond to Ca occupancies between 90-95\%. Right inset: examples of peak splitting fitted by a two-phase model.}
\label{refinement}
\end{figure}

Rietveld refinements of the high-resolution synchrotron powder XRD patterns were carried out in the space group $P6_3/mcm$ (Table \ref{crystaldata}, Table \ref{coordinates} and Figure \ref{refinement}) \cite{ICSD}. Close inspection of the $h00$ and $00l$ peaks revealed anisotropy (which could not be explained by symmetry-lowering) consistent with a two-phase model (see inset of Figure \ref{refinement}). Note that this small shift in lattice parameters could not be seen on a lab diffractometer and requires high-resolution diffraction data.  This second, minority phase occurs in the same space group as the majority phase with slightly different $a$ and $c$ lattice constants and was necessary to accurately model the pattern (inclusion of a second phase decreased the overall $\chi^2$ from approximately 22 to less than 15).  Further systematic peak broadening of the $00l$ reflections was modeled with selective peak broadening in the last stages of the refinement.  

The freely refined site occupancies of Ca and P for the majority and minority phases converged to ratios within error of 3:2 and 5:3, respectively. The majority phase contributes roughly 80\% by weight, whereas the minority phase is approximately 20\%. Based on Corbett and Leon-Escamilla's previous characterization of stuffed hydride Mn$_5$Si$_3$-type compounds, such as Ca$_5$As$_3$H \cite{corbett1998widespread,leon2006hydrogen}, the minority phase was modeled as Ca$_5$P$_3$H. We attribute the presence of interstitial H to hydride impurities in the reactant alkaline earth metal or to H$_2$ diffusion through the Ta tubes at high temperature, both of which are common sources of contamination \cite{corbett1998widespread}. While we acknowledge that x-ray diffraction is expected to be insensitive to H-scattering, the inclusion of H on the $2b$ site is necessary through chemical reasoning to obtain a charge-balanced compound for the minority phase, discussed in further detail below.

The minority phase was refined with an overall thermal parameter, whereas the atoms in the majority phase were treated with isotropic thermal parameters (the Ca sites were constrained to have the same thermal parameter).  In the majority phase, the composition was determined by performing a series of refinements where the occupancies of the Ca sites (constrained to have the same percent occupancy) were fixed at incrementally changed values, and the other parameters were allowed to freely refine.  This accommodates the possible presence of correlations between refinement parameters \cite{mcqueen2009extreme}. As shown in the inset of Figure \ref{refinement}, the $R_{\text{Bragg}}$ for the majority phase reached a minimum when the Ca sites were 90-95\% occupied, yielding a formula within error of Ca$_3$P$_2$.  The occupancies of the Ca sites were therefore fixed to stoichiometric Ca$_3$P$_2$ in the final stages of the refinement.  No evidence of vacancy ordering or P deficiency was found.  The minority phase consistently refined to full occupancy of the Ca and P sites (the occupancy of H was not refined), yielding the charge-balanced compound Ca$_5$P$_3$H.  Additional tests were performed with oxygen as the interstitial ion stabilizing the minority phase, but this led to poorer agreement of the model to the data, and thus that possibility was eliminated.  The final atomic coordinates and occupancies for both phases are given in Table \ref{crystaldata}.  Two impurity phases are also present in small proportions, Ca$_4$P$_2$O (2.5\%) and Ca$_x$TaO$_3$ (0.02\%).  No attempt was made to obtain the minority hydride phase in pure form.

\begin{figure}[htbp]
\centering
\includegraphics[width=12cm]{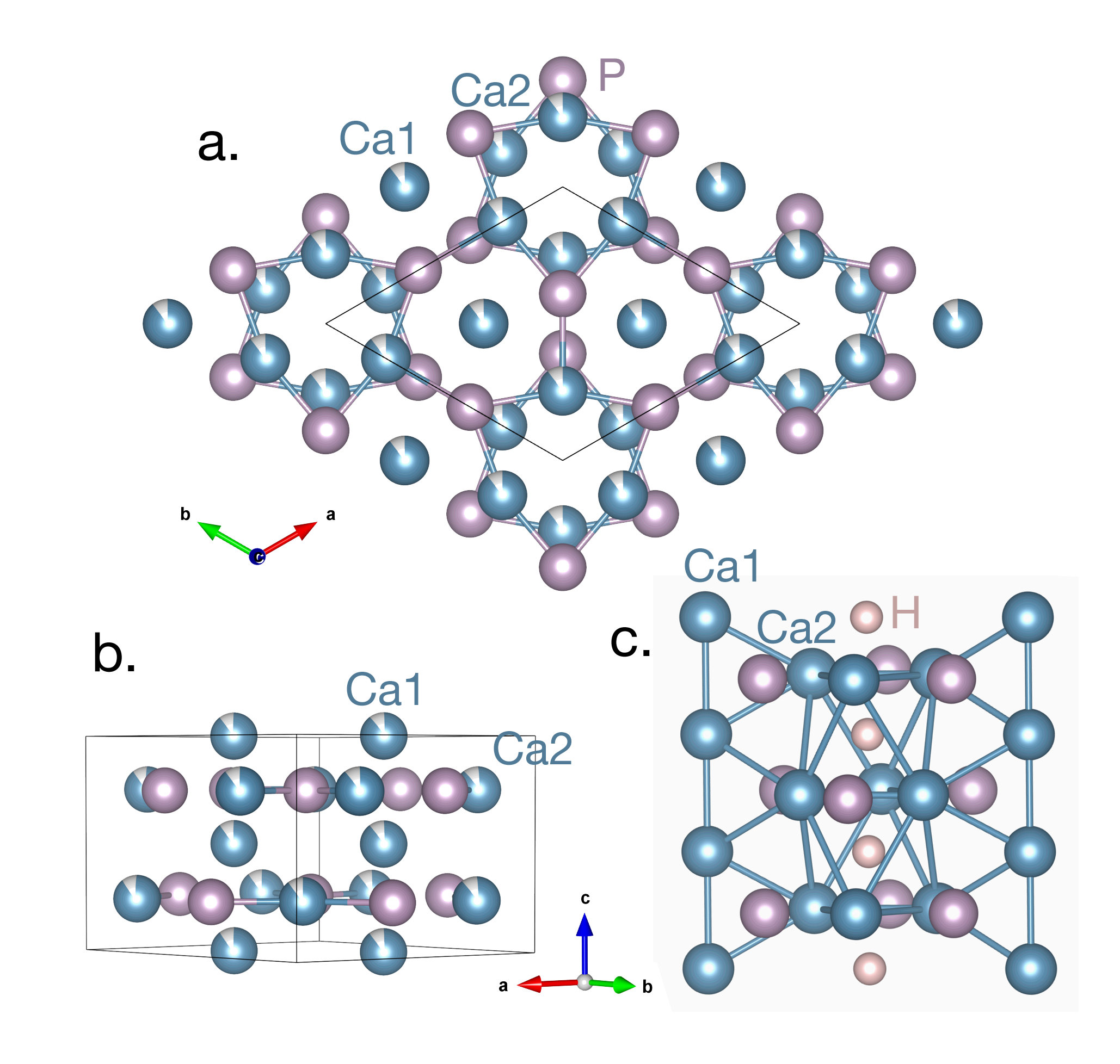}
\caption{The crystal structures of Ca$_3$P$_2$ and Ca$_5$P$_3$H. a. View of Ca$_3$P$_2$ along the [001] direction. b. Unit cell of Ca$_3$P$_2$ with planar network of Ca--P atoms indicated. c. Representation of H interstitial atoms in Ca$_5$P$_3$H in octahedral coordination with Ca atoms.}
\label{structures}
\end{figure}

Visualizations of the structures of Ca$_3$P$_2$ and Ca$_5$P$_3$H are shown in Figure \ref{structures}. The lattice parameters of the phases are similar (Table \ref{crystaldata}), with Ca$_5$P$_3$H having a slightly smaller unit cell volume (0.1\%). This is in line with the volume trends observed by Leon-Escamilla and Corbett for alkaline earth arsenides of the Mn$_5$Si$_3$-type structure in the presence of interstitial hydrogen \cite{leon2006hydrogen}, albeit the volume change here is to a lesser degree. We attribute this to the fact that the Ca deficiency in the hydrogen-free compound reduces the volume of this phase relative to the other alkaline earth pnictides with all sites fully occupied.

As Leon-Escamilla and Corbett have demonstrated, the inclusion of hydrogen in Mn$_5$Si$_3$-type phases serves a chemically straightforward stabilizing function \cite{corbett1998widespread,leon2006hydrogen}. Electron counting reveals that a compound with the formula $A_5B_3$, where $A$ is an alkaline earth metal and $B$ is a pnictogen, has one extra valence electron per formula unit (assuming the oxidation states of +2 for $A$ and -3 for $B$). The inclusion of an interstitial atom such as H or F adopting a -1 oxidation state yields a closed-shell, charge-balanced system. For the light and relatively polarizing elements Ca and P, we expect a metallic Ca$_5$P$_3$ phase to be more unstable than the analogous $A_5B_3$ phases with heavier elements. The above electron counting scheme can be applied to qualitatively rationalize both Ca deficiency in the Ca$_3$P$_2$ phase and the H interstitial in the minority Ca$_5$P$_3$H phase.

Previous thermodynamic studies of the Ca--P chemical system have noted that systematic direct study of the phase behavior is difficult owing to the high volatility and reactivity of the constituents \cite{zaitsev1991thermodynamic}. Based on conditions for Ca$_3$P$_2$ synthesis given in the existing literature \cite{zaitsev1991thermodynamic,pytlewski1963study}, we hypothesize that the Mn$_5$Si$_3$-type compound reported here may be a higher-temperature phase favored by the sealed environment of the tantalum reaction container. Detailed structural characterization of the polymorph reported in previous studies \cite{pytlewski1963study} as well as additional examination of the thermodynamic properties will be necessary to explore the exact relationship between these phases and others in the Ca--P system.

\subsection{Electronic structure of Ca$_3$P$_2$}

\begin{figure}[htbp]
\centering
\includegraphics[width=12cm]{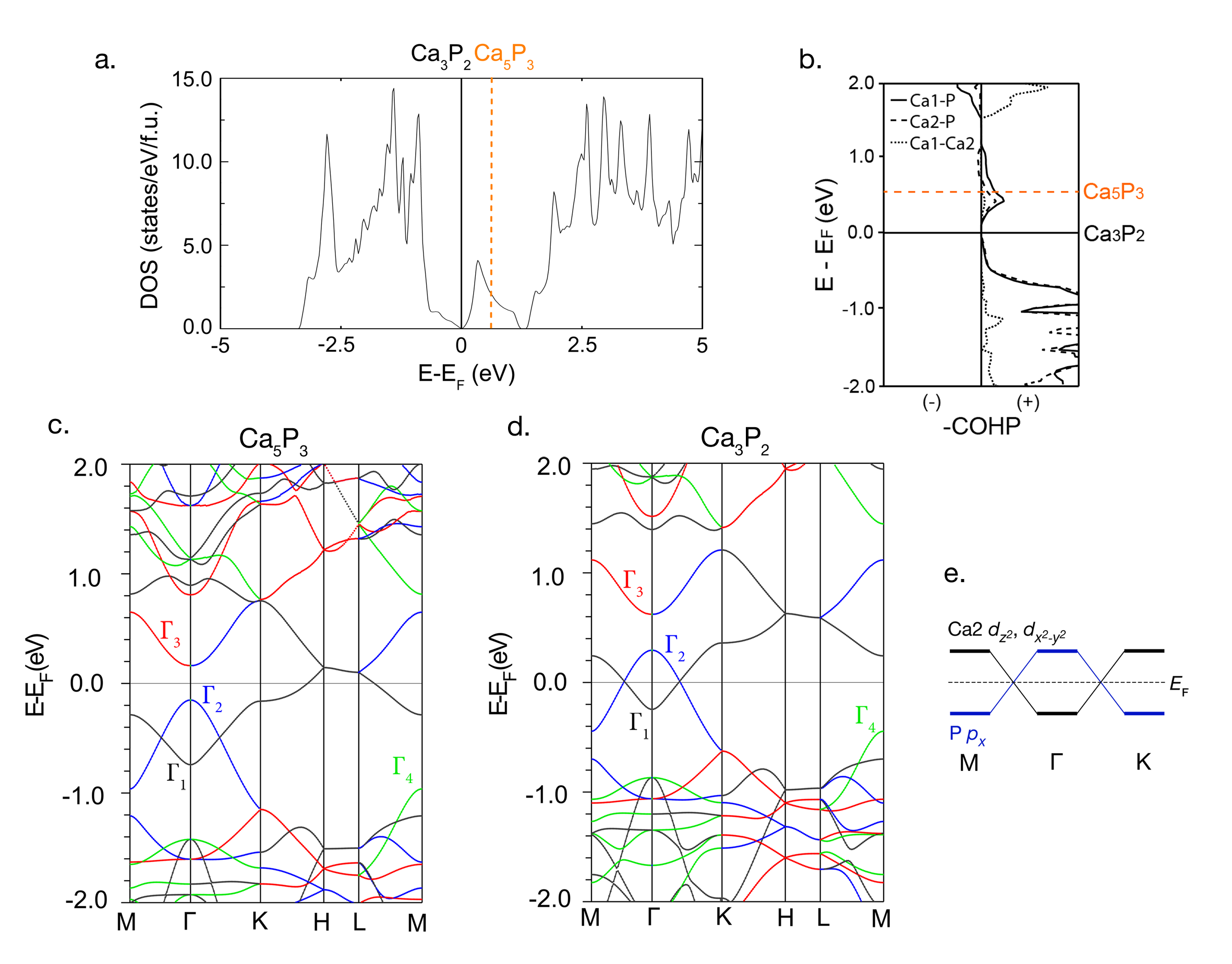}
\caption{a. and b. The density of states (DOS) and the crystal orbital Hamilton population (COHP) of Ca$_3$P$_2$ calculated with the VCA. The Fermi level of hypothetical Ca$_5$P$_3$ (with all Ca sites fully occupied) is indicated. c. and d. The calculated electronic band structures of Ca$_5$P$_3$ (c.) and Ca$_3$P$_2$ (d.), showing bulk Dirac cones between $\Gamma$ and M, and $\Gamma$ and K. Each color corresponds to a different irreducible representation. e. A diagram of the band inversion at the Fermi energy in Ca$_3$P$_2$.}
\label{calculations}
\end{figure}

To further investigate the nature of stability and bonding in Ca$_3$P$_2$, quantitative calculations of the electronic band structure, density of states (DOS), and crystal orbital Hamilton population (COHP) were carried out (Figure \ref{calculations}). Both hypothetical Ca$_5$P$_3$ and Ca$_3$P$_2$ were calculated in order to confirm the validity of the electronic structure model for this system.

Both the COHP and the DOS indicate a region of electron count stability for Ca$_3$P$_2$: the Fermi level lies in a deep pseudogap. The Fermi level of a hypothetical Ca$_5$P$_3$ compound would be located near a peak in the DOS, indicating chemical instability (Figure \ref{calculations}a). Analysis of the COHP indicates that only bonding states are occupied in Ca$_3$P$_2$ (Figure \ref{calculations}b). The additional filled states in Ca$_5$P$_3$ are bonding in nature as well, but the larger COHP value indicates the composition is less stable.

Figure \ref{calculations}c and Figure \ref{calculations}d show the band structure of hypothetical Ca$_5$P$_3$ and Ca$_3$P$_2$ calculated with the VCA. The features of the electronic structures are shifted in energy but otherwise analogous, showing that a rigid band model is valid in this case. In Ca$_3$P$_2$, two linear crossings of bands are observed at the Fermi energy $E_{\text{F}}$ along the the $\Gamma$--K and the $\Gamma$--M lines. In hypothetical Ca$_5$P$_3$, the same feature is present 0.5 eV below $E_{\text{F}}$.

In the $P6_3/mcm$ space group, crystalline symmetry protects these crossings. Along the $\Gamma$--K and $\Gamma$--M lines, the valence band has irreducible representation $\Gamma_1$ and the conduction band has representation $\Gamma_2$. Due to these different irreducible representations, the band crossings are protected by a mirror plane and $C_2$ rotational symmetry in the absence of spin-orbit coupling (SOC). SOC allows mixing of the states and opens the crossing \cite{gibson20143d}. However, the effect of SOC is negligibly small for the light elements Ca and P. Therefore, even when SOC is applied to our calculation, we do not observe a gap in the crossing. 

This type of linear band crossing, also called a bulk Dirac cone, at the Fermi energy has been of recent interest due to interesting physical properties observed in compounds exhibiting such a band structure, such as high magnetoresistance and extremely high mobility \cite{Young2012,Wang2012,Wang2013,borisenko2014experimental,neupane2014observation,Liu2014}. Graphene, for example, is a 2D Dirac semimetal \cite{novoselov2005two,Kane2005Z2,CastroNeto2009}. One of the prominent examples of a 3D Dirac semimetal is Cd$_3$As$_2$, which has recently gained a lot of attention \cite{Wang2013,borisenko2014experimental,neupane2014observation,Ali2014,liu2014stable,liang2014ultrahigh,jeon2014landau}. Although Cd$_3$As$_2$ crystallizes in a very different crystal structure \cite{Ali2014}, the chemical relation to Ca$_3$P$_2$ is still visible. Ca$_3$P$_2$ may be a less toxic alternative to Cd$_3$As$_2$. The other known 3D Dirac semimetal, Na$_3$Bi, has the disadvantage of being extremely air sensitive, which makes handling very difficult \cite{Liu2014}; Ca$_3$P$_2$ is air-sensitive as well, but to a much lesser extent. 

\begin{figure}[htbp]
\centering
\includegraphics[width=12cm]{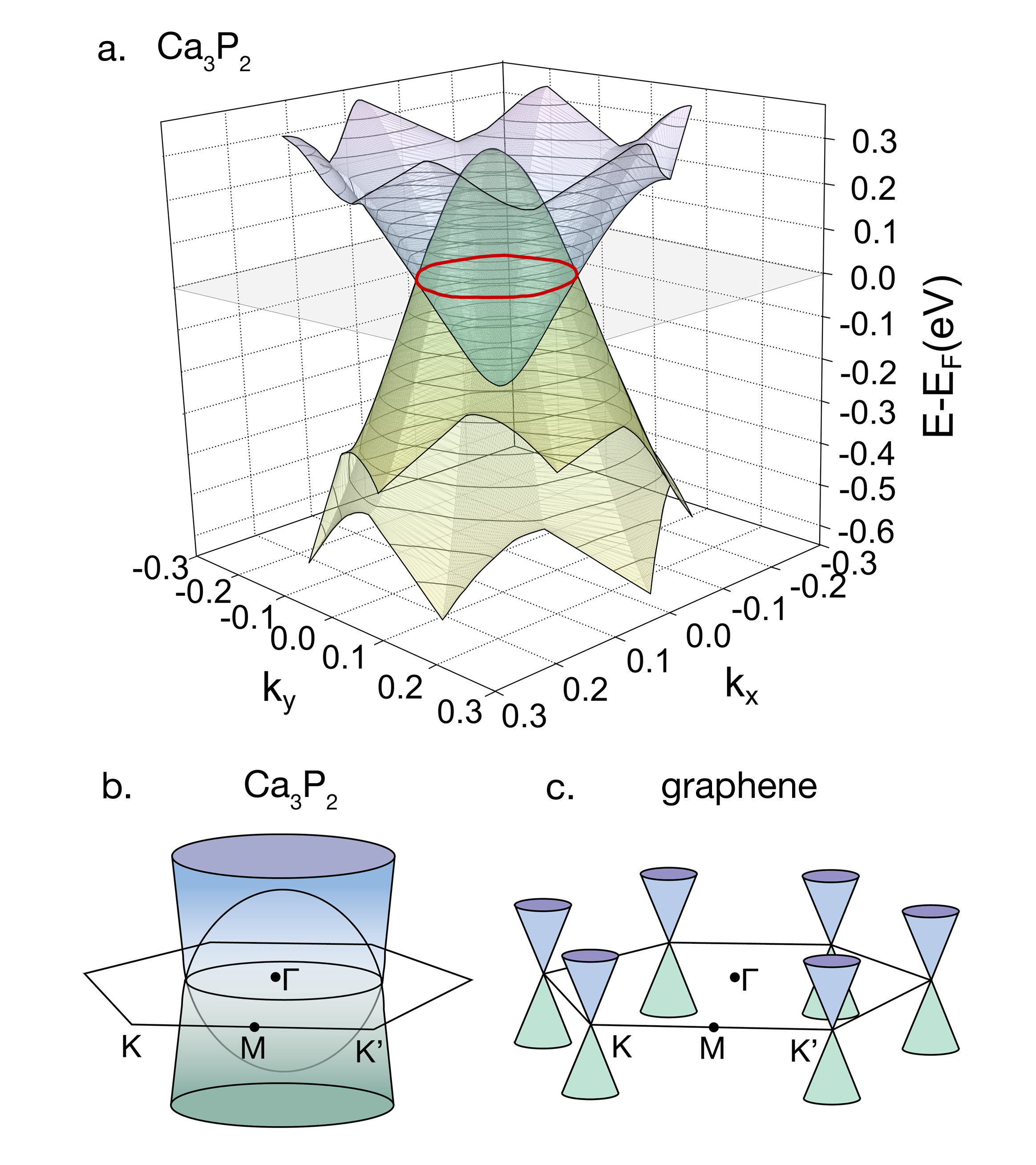}
\caption{a. Three-dimensional plot of the band structure of Ca$_3$P$_2$ along $k_x$ and $k_y$ vs. $E$. The closed ring formed by the band crossings in the $k_x$-$k_y$ plane at the Fermi energy is highlighted in red. b. and c. Graphical representation of the band crossings in the $k_x$-$k_y$ plane of the Brillouin zone for Ca$_3$P$_2$ (b.) and graphene (c.).}
\label{3Dplot}
\end{figure}

Figure \ref{3Dplot} shows a three-dimensional plot of the band structure of Ca$_3$P$_2$ in the $k_x$ and $k_y$ directions vs. $E$. The band crossings with linear dispersions along $\Gamma$--K and $\Gamma$--M form a closed ring around the $C_6$ rotation axis at the Fermi energy. As a result of the C$_{2v}$ symmetry along these lines and the different irreducible representations of the bands, the Dirac crossings are symmetry-protected (without SOC) by the $k_z$ = 0 mirror plane and $C_2$ rotational symmetry. For an arbitrary line connecting $\Gamma$--K and $\Gamma$--M, the point group is $C_s$ and the mirror plane remains a protecting element for all Dirac nodes within the $k_x$-$k_y$ plane. Ca$_3$P$_2$ is therefore the first 3D Dirac semimetal predicted where a mirror plane is the symmetry element that protects the crossing of the bands. Mirror planes have been suggested as a protecting symmetry element by Chiu and Schnyder, who propose a classification scheme for reflection-symmetric Dirac semimetals, among which graphene is an example \cite{chiu2014classification}. An analogy therefore can be drawn between the band structures of graphene and Ca$_3$P$_2$: In graphene, six discrete Dirac points are present at the K and K' points within the $k_x$-$k_y$ plane \cite{novoselov2005two,Kane2005Z2,CastroNeto2009}. In Ca$_3$P$_2$, a continuous, two-dimensional ring of Dirac nodes with sixfold rotational symmetry exists in the $k_x$-$k_y$ plane. Figure \ref{3Dplot}b and c illustrate the relationship.

Additionally, we calculated the band structure of several other compounds in the Mn$_5$Si$_3$ structure type and found the same kind of Dirac crossings 0.5 eV below the Fermi level, similar to hypothetical Ca$_5$P$_3$. The vacancies in Ca$_3$P$_2$ in the Mn$_5$Si$_3$ structure allow for the Dirac states to be at the Fermi level. A hypothetical compound such as Ca$_4$NaP$_3$, for example, should also have Dirac states at the Fermi level assuming a rigid band model. Modulating the electron count of other compounds in this structure type, especially those with light elements, may be a pathway to new 3D Dirac semimetals. Other reported topological materials such as Ru$_2$Sn$_3$ \cite{gibson2014quasi} and [Tl$_4$](Tl$_{1-x}$Sn$_x$)Te$_3$ \cite{arpino2014evidence} also have in-plane mirror planes causing a Dirac crossing if SOC is not applied in the calculation. The large SOC in these materials gaps out these crossings, however. These results imply that in-plane mirror planes can generally produce new 3D Dirac materials if SOC is small enough. 

\section{Conclusions}

The results of the structural refinements suggest that Ca vacancies in a Ca--P compound in the Mn$_5$Si$_3$ structure type yields a composition of Ca$_3$P$_2$, with the driving force being the greater stability of a charge-balanced, closed-shell composition relative to the hypothetical compound Ca$_5$P$_3$. Electronic structure calculations further support the stability of Ca$_3$P$_2$ compared to Ca$_5$P$_3$. The findings suggest that further examination of the equilibrium composition of compounds in the Mn$_5$Si$_3$ structure type, such as other alkaline earth metal phosphides and arsenides, may be of future interest. In addition, study of the exact mechanism and extent of hydrogen interstitials in stuffed Mn$_5$Si$_3$-type structures, drawing upon the existing work of Leon-Escamilla and Corbett \cite{corbett1998widespread,leon2006hydrogen}, may provide additional insight into the composition and stability of these phases and their parent compounds.

The calculated band structure of Ca$_3$P$_2$ suggests the compound may be a candidate for realizing a 3D Dirac semimetal ground state. To the authors' knowledge, this is the first proposed Dirac semimetal in this structure type, as well as the first 3D compound predicted to have bulk Dirac states protected by crystalline reflection symmetry that are negligibly gapped by SOC. The similarities between Ca$_3$P$_2$ and graphene, in particular, merit further study of its electronic properties. The result illustrates that reflection symmetry can operate as a protecting element for realizing 3D Dirac semimetals, which we hope will motivate additional exploration of such systems.

\section{Acknowledgments}

The authors thank Andreas Schnyder for helpful discussion, and 11-BM beamline staff for their assistance. The synthesis and crystal structure analysis was supported by the ARO MURI on Topological Insulators, grant W911NF-12-1-0461, and the electronic structure calculations were supported by the NSF MRSEC program, grant DMR 1420541. Use of the Advanced Photon Source at Argonne National Laboratory was supported by the U. S. Department of Energy, Office of Science, Office of Basic Energy Sciences, under Contract No. DE-AC02-06CH11357.

\section{References}

\bibliography{Ca-P}

\clearpage

\clearpage

\begin{table}[htbp]
\caption{Data of crystallographic refinement for Ca$_3$P$_2$ and Ca$_5$P$_3$H. $\dagger$ Sum of weight fractions is slightly less than 100\% due to the presence of small amounts of impurity phases.}
\begin{center}
\begin{ruledtabular}
\begin{tabular}{r l l}
\toprule
& \multicolumn{2}{c}{Overall refinement} \\
\cmidrule{2-3}
Crystal system  & \multicolumn{2}{l}{Hexagonal} \\
Space group     & \multicolumn{2}{l}{$P6_3/mcm$} \\
Temperature (K) & \multicolumn{2}{l}{295.0} \\
$Z$             & \multicolumn{2}{l}{2} \\
$R_{wp}$        & \multicolumn{2}{l}{25.0} \\
$R_p$           & \multicolumn{2}{l}{23.1} \\
$\chi^2$        & \multicolumn{2}{l}{12.8} \\
\midrule
& \multicolumn{1}{c}{Ca$_3$P$_2$}  & \multicolumn{1}{c}{Ca$_5$P$_3$H} \\
\cmidrule(r){2-2} \cmidrule(l){3-3}
$R_f$                & 11.1        & 11.9 \\
$R_{\text{Bragg}}$   & 12.3        & 13.8 \\
$a$ (\AA)            & 8.25593(1) & 8.25979(2) \\
$c$ (\AA)            & 6.83551(2) & 6.82308(3) \\
$V$ (\AA$^3$)        & 403.491(1) & 403.134(2) \\
Density (g/cm$^3$)   & 2.249      & 2.425 \\ 
Weight fraction$\dagger$ (\%) & 79.1(8)     & 18.4(2) \\
\bottomrule
\end{tabular}
\end{ruledtabular}
\end{center}
\label{crystaldata}
\end{table}

\clearpage

\begin{table}[htbp]
\caption{Atomic coordinates and parameters in the crystal structure of Ca$_3$P$_2$ and Ca$_5$P$_3$H.}
\begin{center}
\begin{ruledtabular}
\begin{tabular}{r l l l l l l }
\toprule
Atom & Site & $x$ & $y$ & $z$ & $B$ & Occ. \\
\midrule
\multicolumn{5}{l}{Phase: Ca$_3$P$_2$} & $B_{\text{iso}}$ & \\
Ca1 & $4d$ & 1/3       & 2/3 & 0   & 1.62(3) & 1 \\
Ca2 & $6g$ & 0.2544(2) & 0   & 1/4 & 1.62(3) & 0.9 \\
P   & $6g$ & 0.6096(2) & 0   & 1/4 & 0.56(3) & 0.9 \\
\multicolumn{5}{l}{Phase: Ca$_5$P$_3$H} & $B_{\text{ov}}$ & \\
Ca1 & $4d$ & 1/3       & 2/3 & 0   & \multirow{4}{*}{2.2(4)} & 1 \\
Ca2 & $6g$ & 0.2543(3) & 0   & 1/4 &  & 1 \\
P   & $6g$ & 0.6104(1) & 0   & 1/4 &  & 1 \\
H   & $2b$ & 0         & 0   & 0   &  & 1 \\
\bottomrule
\end{tabular}
\end{ruledtabular}
\end{center}
\label{coordinates}
\end{table}

\clearpage

\clearpage

\clearpage

\end{document}